\documentclass[]{article}
\usepackage{authblk}
\usepackage{graphicx}
\usepackage[utf8]{inputenc}
\usepackage[english]{babel}
\usepackage{amsmath}
\usepackage{amsfonts}
\usepackage{amssymb}

\usepackage{geometry}
\usepackage{float}

\title{State-dependent driving: A  route to non-equilibrium stationary states }

\author[$1$]{Soumen Das}
\author[$1$]{Shankar Ghosh}
\author[$2,3$]{Shamik Gupta}

\affil[1]{\textit{Department of Condensed Matter Physics and Materials Science, Tata Institute of Fundamental Research, Mumbai 400005, India}}

\affil[2]{\textit{Department of Physics, Ramakrishna Mission Vivekananda
    Educational and Research Institute, Belur Math, Howrah 711202, India} }
\affil[3]{\textit{Regular Associate, Quantitative Life Sciences Section, ICTP-The Abdus Salam International Centre for Theoretical Physics, Strada Costiera 11, 34151 Trieste, Italy}}

\date{}
\begin{document}

\maketitle

\begin{abstract}
 We study three different experiments that involve  dry friction and periodic driving, and which employ both single and many-particle systems. These experimental set-ups, besides providing a playground for investigation of frictional effects, are relevant in broad areas of science and engineering. Across all these experiments, we monitor the dynamics of objects placed on a substrate that is being moved in a horizontal manner.  The  driving couples to the degrees of freedom of the substrate,	and this coupling in turn influences the motion of the objects.  Our experimental findings suggest emergence of  stationary-states with non-trivial features. We invoke a minimalistic phenomenological model to explain our experimental findings.  Within our model, we  treat the injection of energy into the system to be dependent on its dynamical state, whereby energy injection is allowed only when the system is in its suitable-friction state. Our phenomenological model is built on the fact that such a state-dependent driving results in a force that  repeatedly toggles  the frictional states in time, and serves to  explain our experimental findings.

\end{abstract}

\section{Introduction}

A common parlor trick to demonstrate inertia is to place a coin on a
post card, and then quickly pull the card. If the pull is quick enough,
the coin remains almost in its place while the card gets pulled out.
While the trick and its many variations (pulling a book out of a
stack of books, or, the game Jenga) are themselves very instructive,
they nevertheless leave a few pertinent questions related to the
stationary-state dynamics unanswered. The coin is attached to the card by	forces of friction. Common wisdom tells us that if the acceleration of	the card is larger than a critical value, the coin will begin to slip	with respect to the card.  However, as soon as the coin starts to slip, a crack opens up on the interface~\cite{malthe2021onset,svetlizky2014classical}, and this results in weakening of the  frictional coupling between the card and the coin. Since there is dissipation, the coin loses momentum and eventually heals the  crack, and the frictional coupling regains its strength. The whole process of weakening and strengthening of the frictional coupling  undergoes repeated cycles in time. In such a scenario, does the coin eventually move with uniform velocity or does it accelerate? In this paper, we will show that  the qualitative aspects of this particularly simple experiment and some variants of it can be explained in terms of a phenomenological model in which the rate of energy injection into a system depends on its dynamical state.

\subsection{Degrees of freedom and internal states of a system \label{sec_dof}}
In a more general setting, the aforementioned scenario would be an
example of an interacting system evolving in presence of drive and
dissipation. In our studied systems, input of energy through the external drive at the level of the substrate (e.g., the card)   affects the evolution of the frictional states of the objects (e.g., the coin) placed on the substrate and may cause in time random transitions between the different frictional states. These transitions in turn would act back on the time evolution of the objects, resulting in an intricate  (and intriguing, in many-body complex systems) interplay between the dynamics of the objects and that of the frictional states  ~\cite{kumar2015granular,das2020tuning}. In the context of the coin and the plate, the  qualitatively different frictional states are that of slipping (low-friction state) and sticking (high-friction state).   If the coin is replaced by a  sphere, then the frictional  states may be identified with rolling (low-friction state) and sliding (high-friction state) motion.

It is evidently of interest to ask: what is the long-time behavior of the aforementioned class of systems? Does the system achieve a stationary state with a time-independent distribution of the degrees of freedom of the objects? We may in general anticipate that a balance between drive and dissipation does in fact lead the system to a stationary state. Presence of an external drive precludes the possibility for the stationary state to be an equilibrium one. Consequently, any stationary state the system relaxes to at long times would be a generic nonequilibrium stationary state (NESS) ~\cite{livi2017nonequilibrium}. In this work, we highlight the subtleties that come into play in the dynamics,  by presenting three experiments - (i) a single coin on a platform oscillating along one axis and (ii) a single sphere
and (iii) a collection of spheres placed on an orbiting platform ~\cite{aumaitre_granular_2003}. The first experiment is one dimensional in nature, and from its detailed study, we  establish that  at  a phenomenological  level, the coupling between the plate and the coin can be modeled by a viscous force-like velocity dependent term. In the second experiment, the center-of-mass of the sphere covers a two-dimensional space, and it does so under the influence of two  orthogonal forces. From this  experiment, we establish that the stochasticity in the trajectory of the sphere comes from the  randomness in the amplitudes of the two orthogonal forces. In the third experiment, we introduce  interactions between the particles, and establish that the  resulting velocity distribution attains  a unique  velocity distribution with  the tail, which is exponential in nature,  gaining weight with increasing density of spheres. This observation is  surprising, since collision-induced cooling is at the heart of many observable NESS states in granular systems.

Following the experimental results, we present a minimalistic one-particle theoretical model  that  captures the essence of the experimental findings, and  provides us with a framework to understand such scenarios. Within our model, we treat the injection of energy to the system to be dependent on  its  dynamical state,  whereby  energy  injection  is allowed  only  when  the  system  is  in  a suitable-friction state of the system. The idea of state-dependent driving is rooted in various biological systems, where the functionality of a state is maintained by an organism by constantly regulating  the activity of its  various internal processes.  In the context of active systems,  this would translate to the idea of prescribing a  local density-dependent mobility to the particles.  It is known that such  quorum-sensing interactions can give rise to motility-induced phase transitions~\cite{tailleur2008statistical}.  Another example is that of enzymes locally enhancing diffusion  by self-regulating  the  phoretic ~\cite{golestanian2008mechanical} and hydrodynamic forces generated by themselves. This can lead to such functional behavior as antichemotaxis ~\cite{jee2018enzyme}.

\subsection{Dry friction models and their limitations \label{sec_dry_friction}}
The experiments and numerics  presented in this paper are  considered for  frictional systems.  At a first glance, the   experimental scenarios considered here might appear  commonplace and well studied.  It is thus important that we discuss the  conventional method of treating  apparently  similar  problems using  the   well-established framework of  friction and  point out the differences between the ones that are exhaustively discussed in the literature from the ones  studied here.

\paragraph{\textbf{The  problem with dry friction:}}
A  solid  that is frictionally coupled to a substrate is put to motion when the force  $F$  applied to the body exceeds $\mu \vert N \vert $, where $ N $ is the normal force exerted  by the solid on the substrate and $\mu$ is the coefficient of friction. When the relative velocity between the substrate and the body is zero, then $\mu=\mu_s$, the coefficient of static friction, and  $ \vert F_F \vert \le \mu_s N $, where $F_F$  is the force of friction. However, when the object is in motion, then $\mu=\mu_k$, the coefficient of kinetic friction, and $ F_F =- \mu_k N$. Here $ \mu_s $ is  a positive real number that characterizes the transition from a stuck to a moving phase,  and $ \mu_k $ is also a positive real number that  characterizes the  friction forces when the  object is slipping with respect to the substrate. The above  description is the Coulomb description of  dry friction. For other phenomenological models of  dry  friction, see Ref.~\cite{pennestri2016review}.  In general, the coefficients of friction are  defined  for a pair of materials,  and  $ \mu_s \ge \mu_k $. Thus, $ F_F $ forms an admissible set $ \mathbb{C} $, i.e., $\mathbb{C}=\{F_F\vert-\mu_s N\le F_F \le \mu_s N\} $. Unlike viscous  friction, where $ F_F \rightarrow 0 $  as $ v\rightarrow 0 $, there exists a velocity gap $ \Delta v $ in dry friction, i.e., the velocity becomes zero if the  magnitude of the applied force $ F $  becomes smaller than a threshold value ~\cite{ghosh2017geometric}.  Dry friction breaks the time-reversible symmetry associated  with the dynamics  of bodies. This loss of symmetry is sometimes reflected in the  non-linear (often chaotic)  dynamics exhibited by these systems ~\cite{stelter1992nonlinear,denny2004stick}.  The presence of an admissible set of  friction values and  the velocity gap  makes the problem of dry friction not amenable to an analysis within the framework of deterministic dynamics.

\paragraph{\textbf{Modeling  Friction as a mechanical instability:}}
Friction is considered to be a nonlinear phenomenon ~\cite{urbakh2004nonlinear}, and models like the Prandtl-Tomlinson   associate this nonlinearity  with  mechanical instabilities (for an historical account, see Ref. ~\cite{schwarz2016exploring,popov2014prandtl}).  Within the framework of these models, the overall response of  the mechanical system is understood in terms of  the ratio of two potentials: one  associated with the  substrate - body interaction and the other associated with  elastic coupling between the body  and an external  \textit{anchor point} ~\cite{vanossi2013colloquium,persson2013sliding}. This \textit{anchor point} can either   remain stationary or  have a well-defined trajectory. If  the ratio of these two potentials is less than one, the overall motion of the body is smooth. On the other hand, when this ratio is large, the sliding motion becomes jagged (slip-stick motion), and the system  continuously exhibits a  series of \textit{elastic instabillity} transitions (see chapter 10 of Ref.~\cite{persson2013sliding}).  Thus, Prandtl-Tomlinson is essentially a two-body  model which  in itself or whose close variants like the ones that incorporate stress-augmented thermal activation process~\cite{krylov2005thermally} has been  successfully used to understand a wide range of frictional problems encountered in experimental studies  using a variety of friction  force microscopes (FFM) ~\cite{popov2014prandtl,schwarz2016exploring,gnecco2015fundamentals}. The dynamics of the model that comprises a single particle being dragged with velocity $v$ on a one-dimensional substrate considered to be a  sinusoidal potential is described by the equation of motion
\begin{equation}
    m\partial_t v=-\gamma v +F(x,t)+\eta(t),
\end{equation}
where $\eta(t)$ represents Gaussian, white noise satisfying $\langle \eta(t)\rangle=0;~\langle \eta(t)\eta(t')\rangle=2\gamma k_BT\delta(t-t')$ and $T$ is the temperature. In the above equation, $\gamma$ is the damping coefficient, while the force $F(x,t)=-\partial_x U(x,t)$ is obtained from the potential $U(x,t)=U_0\cos(2\pi x/a)+(K/2)(x-vt)^2$. Here, $U_0$ and $a$ are respectively the amplitude and the periodicity of the sinusoidal potential, while
$K$ represents the strength of the harmonic coupling between the particle and the substrate.
Friction associated with  extended objects is treated   within the  framework of Frenkel-Kontorova (FK) like models ~\cite{braun2004frenkel}. Here, the  elastic energy associated with the  coupling between the body and  the \textit{anchor point} is replaced by   the energy associated with the deformation of the body itself.  As a result, the response of soft objects  in presence of frictional force is very different from that of the  more rigid ones - a phenomenon that is  known as  the Aubry transition ~\cite{peyrard1983critical}. Rigid objects tend to move smoothly, while soft objects exhibit slip-stick motion.

Almost all experimental realizations in friction are classified into two types, one where the  force  is directly applied to the  object, e.g., an atomic force microscope tip  being dragged over a substrate  or where the body is held with  a spring  to a point in space and  the  substrate   moves below it ~\cite{sandor2013chaos,abe2014intermittency,carlson1989properties, gagnon2020review}. In either case, the two competing energy scales can be easily identified and hence, these situations are amenable to detailed analysis.  However,  there may be situations, e.g. coin on a  horizontally vibrating plate ~\cite{buguin2006motions}, where identifying two separate energies, one of which is due to the application of a force on the system of interest and the other due to its motion against a frictionally-coupled substrate, may not be possible. This happens when the  processes by which energy is injected  and dissipated are not separated in space and time. This is precisely the dynamical scenario that we address in this work.

\paragraph{\textbf{Stochastic aspects of Friction:}}
Though the force of friction is an easily measured quantity,   its value is  highly sensitive to  experimental conditions. Often, measured friction forces vary from one experiment to another and also these are known to strongly age with time ~\cite{muser2008static,li2011frictional}.  For dry friction, the local structure of the interlocking interfaces plays a key role. Since the area of this  interlocking region is small,  each  patch of the interface corresponds to different values of friction. This lack of self-averaging necessitates the use of a stochastic dynamics to study the problem. In doing so, one may invoke stochasticity in the configuration space through the modeling of  making and breaking of contacts ~\cite{albertini2021stochastic,feng2003discrete}.  Alternatively, one may invoke stochasticity in the energy space whereby energy injection and dissipation are modeled as stochastic processes.  In this paper, in contrast to earlier work, we  follow the second approach and propose a model in which the energy transfer between a moving substrate and a body is  treated as a stochastic variable whose value depends on the \textit{`state'} of the system. Here, we use the term \textit{`state'} in the same sense as described in the leading paragraph of Sec. \ref{sec_dof}.  One may recall that phenomenological models  of friction like the `rate-state model'  ~\cite{dieterich1979modeling,ruina1983slip,van2018comparison,tian2018rate} also invoke the concept of states in a different sense,  without however  explicitly identifying the states. However,  many  recent works have attempted to do so by treating friction as localized  electro-plastic response of the contact zone~\cite{baumberger1999physical}.


\begin{figure}[t]
    \centering
    \includegraphics[width=.7\linewidth]{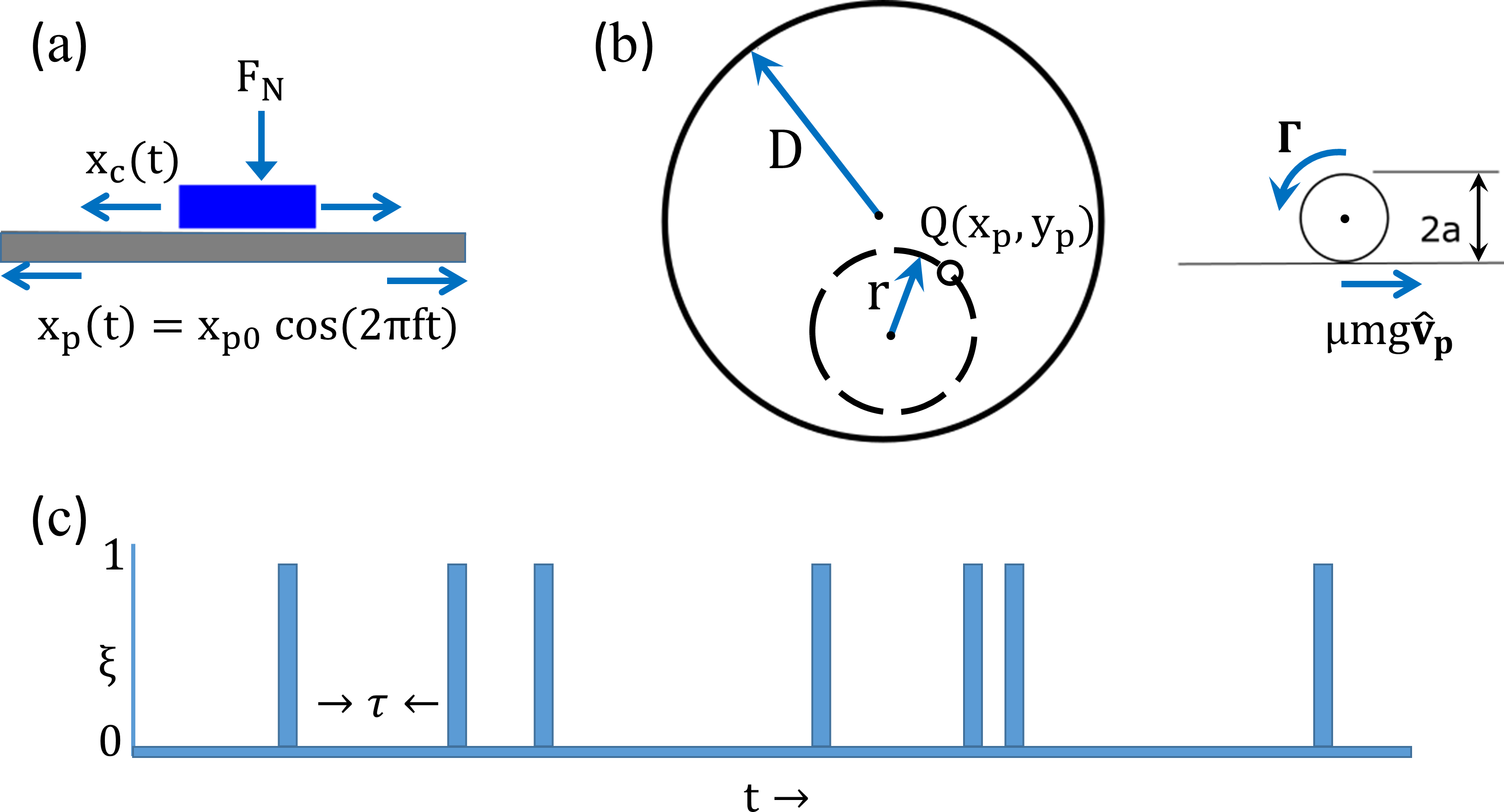}
    \caption{(a) First experimental setup: a coin on a plate that is
        being vibrated sinusoidally in time. (b) Second
        experimental setup: Stainless steel balls of diameter $2a=800 \,\mathrm{\mu
            m}$ confined in a cylindrical container of diameter
        $D=150 \,\mathrm{mm}$ and height $1 \,\mathrm{mm}$, with a glass
        plate used as a lid for the container to enforce a
        two-dimensional geometry for the problem. The container is
        placed on a square platform, and the entire assembly along with
        a camera is made to perform an
        orientation-preserving horizontal circular motion. The net
        motion of the assembly  may be imagined as a combination of two
        circular rigid-body motion of the platform: (i) a rotation about
        shaft 1 with angular frequency $\mathbf{\omega}$, and (ii) an
        opposing rotation with angular frequency $-\mathbf{\omega}$
        about shaft 2 that passes through the center of the platform.
        (c) Typical temporal variation of the dichotomous noise $\xi(t)$
        used in the phenomenological dynamics~(\ref{Eqn:A} and \ref{Eqn:B}) to model
        both the experiments.}
    \label{Fig1}
\end{figure}

\section{Results}
\subsection{A coin on an oscillating plate}

\begin{figure}[t]
    \centering
    \includegraphics[width=0.9\linewidth]{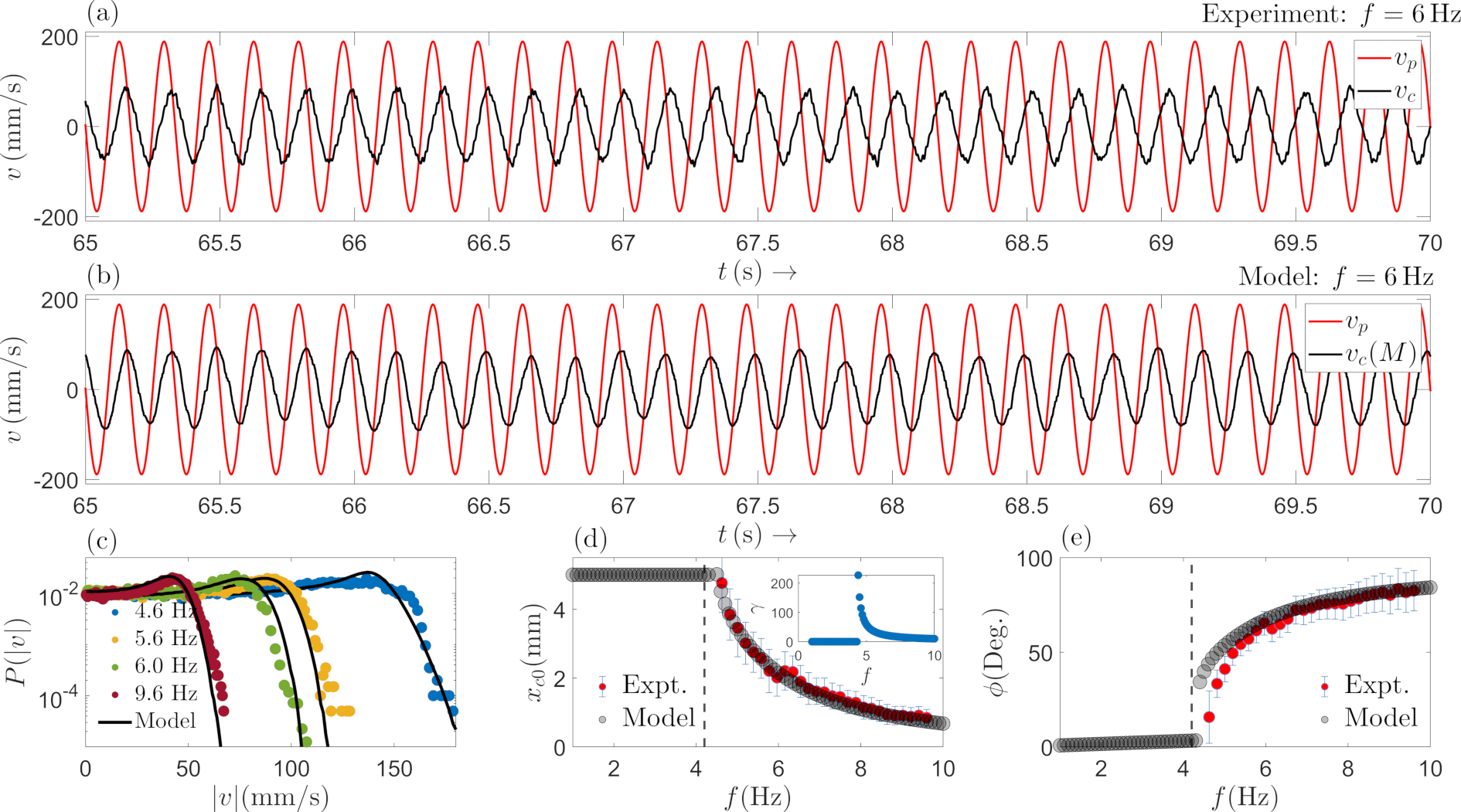}
    \caption{A coin on an oscillating plate:  The trace of the velocity of the coin (black) and a point fixed on the plate (red) at driving frequency $f=6$ Hz are both shown in panel (a).  Panel (b) shows results from numerical simulation of the phenomenological model (Eqn.~\ref{Eqn:A}).
        The probability distribution of the magnitude of the velocity of the coin in the laboratory frame is plotted in panel (c) at the representative values of $f$. The filled symbols correspond to that obtained from the experiment while the solid lines correspond to that obtained from the model. The  time-averaged amplitude of the displacement of the coin ($x_{\rm c0}$) and the  phase difference $\phi$ of the coin with respect to the drive is shown in panel (d)  and (e), respectively,  as a function of the driving frequency $f$. The red symbols correspond to that obtained from the experiment while the grey symbols correspond to that obtained  from the model. The dotted line in (d) and (e)   separates the stuck phase ($f\le f_c$) from the  slipping phase ($f>f_c$). 	Here $f_c=4.4$ $\rm{Hz}$ is the critical frequency below which the coin is completely stuck to the plate. 	For the simulation, for $f>f_c$ we took $\gamma \tau_p/m_p \sim 12$, $\beta^{-1} \sim (f-f_c) $, and $\gamma$ was taken to  vary inversely with  $f-f_c$ (see  inset of (d) and appendix  for additional information on the parameters) .
    }
    \label{Fig2}
\end{figure}

A coin is placed on a plate that is being vibrated in the $x$
direction, see Fig.~\ref{Fig1}(a). The motion of both the coin and the plate is measured
simultaneously. The plate moves sinusoidally, $x_{\rm p}(t)=x_{\rm p0}\cos(2
\pi f t)$, where $x_{\rm p}(t)$ is the position of a point on
the plate, $x_{\rm p0}$ is the amplitude of vibration, and $f$ is the
driving frequency.

\subsubsection{Experimental results}
For small values of $f(\le f_c=4.4$ $\rm{Hz})$, the coin remains stuck to the
plate, while for larger values of $f$, it begins to slip with respect to
the plate. Typical time traces of the velocity of the center of mass of
the coin and a point fixed on the plate for $f=6$ Hz are shown in
Fig.~\ref{Fig2}(a). In the phase in which the coin slips with respect to
the plate, its center of mass drifts slowly in time. However, for small
time intervals, its response $x_{\rm c}(t)$, namely, the location of its
center of mass, is approximately sinusoidal, and has the
same frequency of variation as that of the plate. The probability distribution of $|v|$,
the magnitude of the velocity of the center of mass of the coin, is
plotted in  Fig.~\ref{Fig2}(c); the distribution is obtained by sampling the trajectories in the long-time limit over a time
window. The variation of the time-averaged amplitude of displacement of the coin  and its  phase difference $\phi$  with respect to the plate  in the long-time limit is plotted in
Fig.~\ref{Fig2}, panels (d) and (e), respectively,  as a function of the driving frequency.

Our main	experimental  observations may be summarized as follows:
\begin{enumerate}
    \item At long times, the velocity as well as the displacement of the coin is approximately sinusoidal
    in time (Fig.~\ref{Fig2}(a)).
    \item  The velocity distribution of the coin has a tail that is exponential (Fig.~\ref{Fig2}(c)).
    \item  The time-averaged amplitude of the displacement of the coin $x_{\rm c0}$ decreases with the drive frequency  (Fig.~\ref{Fig2}(d)).
    \item  The phase difference $\phi$  saturates to a large value with increase in the drive frequency (Fig.~\ref{Fig2}(e)). This is in contrast to what is seen in the case of a damped, driven harmonic oscillator with a real damping constant, which in turn precludes use of noisy harmonic oscillator to model our experimental scenario. The case of a complex damping constant is discussed in the appendix.\\
\end{enumerate}

\subsubsection{Discussion}

The most evident way to describe the system would be to use the Coulomb friction model ~~\cite{persson2013sliding}.  If  the  driving frequency is such that the amplitude of the sinusoidal forcing exerted by the  plate is larger than the maximum static friction force, i.e., $|m(2\pi f)^2x_{\rm p0}|>\mu_s mg$, with $\mu_s$ being the coefficient of  static friction between the coin and the plate and $m$ the mass of the coin, the coin will undergo both stick and slip motion periodically.  In the stuck state, i.e., when $|m(2\pi f)^2x_{\rm p0}\cos(2\pi ft)|<\mu_s mg$, trajectory of the coin will follow the sinusoidal nature of the trajectory of the plate. However, as soon as the driving force exceeds $\mu_s mg$, the coin will start to slip with respect to the plate and it will then move under a constant acceleration due to the  force of kinetic friction $\mu_k mg$. Here,  $\mu_k$  is the coefficient of  kinetic friction between the coin and the plate.	Hence, the velocity of the coin will be partly sinusoidal and partly linear in a single period. This model gives rise to a deterministic velocity profile. This is in stark contrast to the noisy sinusoidal nature of the velocity observed in the experiments. In addition, the  velocity distribution, the amplitude and the phase response of the coin as obtained from the Coulomb friction model of friction is very different from what is observed in the experiments. We have provided the details in the appendix.

\subsubsection{Model}

The response of the system   for $f>f_c$ and $f<f_c$ is qualitatively different. For $f<f_c$, the coin and the plate  behave as a single rigid object, i.e., $x_c(t) =x_p(t)$, which is unlike the situation for $f>f_c$.  The description below is   limited to the latter case only.

Experiments suggest that the coin never gets stuck to the plate at the smallest time scale over which the motion of the coin and the plate is measured, i.e., $5.5$ $\rm{ms}$. Hence, to explain the experimental observations, we propose that the coin has access to (i) a `high-friction' state (which in a  limiting case becomes the  stuck state) and (ii) a `low-friction' state (or a slip state). Without the presence of `high-friction' state, the coin would never experience the external driving applied to the plate and hence, the response of the coin would never reflect the waveform of the drive.  However, over a large parameter regime, we find that the amplitude of the coin is  smaller than that of the plate. Thus, it is reasonable to assume that the `high-friction' state  survives over a timescale that is smaller  than the  smallest time scale of measurement, and that during this time scale, $\tau_{\rm p}$, a fraction of the momentum of the plate gets transferred to the coin. Here, it becomes necessary to differentiate between a `stuck' state (observed for   $f< f_c$) and a `high-friction' state. In contrast to a `high-friction' state, in a `stuck' state, the entire momentum and hence, the entire force would be transferred from the plate to the coin. We assume the transitions between the high-friction and the low-friction state to occur randomly in time. We thus propose a stochastic variant of the Coulomb friction model for the motion of the coin	on the vibrated plate which captures the essential qualitative	features of its dynamics observed in the experiments described above.

To	be consistent with the experiment, we choose to describe the motion of the center of mass of the coin in an inertial frame, i.e., in the laboratory frame. With respect to such a frame, the	plate on which the coin is placed is being vibrated sinusoidally in	time, so that its velocity at time $t$ reads $v_{\rm p}(t)=v_{\rm p0} \sin (2\pi f t)$. When the coin is in the `high-friction' state, the equation of motion of its center of mass is the usual Newton’s equation of motion:
\begin{equation}
    m \partial_t v = F(t);~F(t) \equiv m_{\rm p} v_{\rm p}(t)/\tau_{\rm p},
\end{equation}
where $v$ is the velocity of the coin, while $F(t)$
is the force experienced by the coin due to transfer of momentum from
the plate to the coin. Here, the quantity $m_{\rm p}$
is the mass of the plate and $\tau_{\rm p}$ as defined earlier is the timescale over which the `high-friction' state survives and during which the momentum of the vibrating plate is transferred to the coin.
On the other hand, when the
coin is in the `low-friction' (slip) state, it is detached from the plate, and
consequently, it moves in presence of an effective damping force that
dissipates in time the initial momentum of the coin at the instant of
detachment from the plate. Hence, in the `low-friction' state, the equation of motion of center of mass is given by
$m \partial_t v=-\gamma (v-v_p)$,
where $\gamma>0$ is a phenomenological dissipation constant. The coin toggles randomly in time between
the `high-friction' and the `low-friction' state. Introducing a random variable $\xi$
taking on values 1 or 0 corresponding to the `high-friction' and the `low-friction' state, respectively, one may on the basis of the foregoing write  down the equation of motion of the center of mass as a stochastic differential equation, a Langevin-like equation, of the form
\begin{equation}
    m \partial_tv=-\gamma (v-v_p)+\xi(t)(F(t)+\gamma (v-v_p)).
    \label{Eqn:A}
\end{equation}

Equation~(\ref{Eqn:A}) has to be supplemented by another equation describing
the time evolution of the instantaneous $\xi$, namely, $\xi(t)$. While derivation of an exact form of the latter would invariably
involve a detailed modeling of the dynamics of the contact region via which the friction force is transmitted, we here offer a
phenomenological description for the evolution of $\xi$ in terms of a
stochastic Markov process, namely, between times $t$ and $t+{\rm dt}$,
the variable $\xi(t)$ is updated to read $\xi(t+{\rm d}t)=1$ with
probability $\beta {\rm d}t$, while $\xi(t+{\rm d}t)=0$ with the
complementary probability $1-\beta {\rm d}t$. Here, $\beta>0$ is a
dynamical parameter. It then follows that the
random time $\tau$ between two successive occurrences of the value
unity for $\xi$ is distributed as an exponential: $p(\tau)=\beta
e^{-\beta \tau};~\tau \in [0,\infty)$, and that the average $\tau$ is given
by $\langle \tau \rangle=1/\beta$. As a function of time, $\xi(t)$
appears as a set of impulses distributed randomly in time, as shown in
Fig.~\ref{Fig1}(c).  For a given drive, the parameter $\beta$ is a phenomenological parameter.
To fit our experimental data we require $\gamma \propto 1/(f-f_c)$. This suggests 	strain rate  induced weakening  of the  contact zone   which could possibly arise from the reduction of contact area with increasing  strain rate ~~\cite{scholz2019mechanics}. A reduction in $\gamma$ results in re-scaling of the relaxation time $\tau_p \sim m/\gamma \propto (f-f_c)$, which thereby influences the momentum transfer process via  $F(t)= m_p v_p/\tau_p $.   For the coin to slip, the contact zone must reach a critical size ~~\cite{uenishi2003universal}. 	Further,  given $\tau_p$  describes the  dynamics at the contact  it also likely to influence the time scale associated with the growth of the slip region of the contact zone. We thus assume $\beta^{-1}$ to be a function of $\tau_p$.  This forms the rationale for  assuming a linear dependence between  $1/\beta$ and $(f-f_c)$ which is used  to fit the  data. The specific forms of $\gamma$ and $\beta$ as a function of $f$ are given in the appendix.

It is interesting and pertinent that we draw a parallel between
Eq.~(\ref{Eqn:A}) and the usual form of the Langevin equation that one
encounters in describing say the paradigmatic Brownian motion. Besides
the nature of the stochastic noise, which in the latter is a Gaussian,
white noise and in Eq.~(\ref{Eqn:A}) is a dichotomous noise, one very
important difference is the following. In the case of the Brownian
motion, the strength of the noise term is a constant that does not
depend on the value of the dynamical variable in question (more
precisely, the constant is by virtue of the fluctuation-dissipation
theorem related to the equilibrium temperature of the ambient medium).
By contrast, in  Eq.~(\ref{Eqn:A}), the strength of the noise term is
explicitly dependent on the value of the dynamical variable $v$ being
studied. Another point worth mentioning is that the noise in the case of
Brownian motion is considered stationary, while in our case, the noise
is non-stationary.

Another interesting parallel that one may draw is between the dynamics~(\ref{Eqn:A}) and the dynamics of a driven, damped harmonic oscillator evolving in presence of noise. In such an approach, one may consider the noise to be multiplicative, in the sense that it appears as a coefficient of the damping term in the dynamics, i.e., $m \partial_tv=-\gamma \xi(t)(v-v_p)+F(t)$. The problem with this approach is that even when the coin is slipping, there will be the unphysical dynamics of momentum transfer between the coin and the plate. In contrast, the model presented here is nevertheless founded on physical reasoning, as presented in the paragraph preceding Eq.~(\ref{Eqn:A}).
Note that in the dynamics described by Eq.~(\ref{Eqn:A}), the force
experienced by the coin is being continually reset in time between the
pure sinusoidal drive and the entirely dissipative form as the random
variable $\xi$ toggles in time between its two-possible values.
Equation~(\ref{Eqn:A}) may be considered representative of a class of stochastic dynamical systems in which the force  resets stochastically in time between a set of possible values, the two-state process considered here being the simplest one may conceive. \\

The dynamics described by Eq.~(\ref{Eqn:A}) involves two time scales
$1/\beta$, setting the average time between two impulses, and
$1/f$. We work in the regime in which $1/\beta \ll 1/f$. Consequently,
in the long-time limit, we expect the velocity $v$ to vary sinusoidally
in time with occasional jump in values as an effect of toggling acting
on a time scale that is faster compared to the sinusoidal variation.
As far as the parameter $\gamma$ is concerned, its magnitude would set
the cut-off scale of the velocity in the slip state.\\

Figure~\ref{Fig2}, panels (b)-(e)  compare the results obtained from numerical
simulation of the dynamics~(\ref{Eqn:A}) with the experimental findings. In panel (b), we see that	consistent with our expectations mentioned above, the velocity does
represent an almost sinusoidal variation in time and correspondingly,
the velocity distribution in (c) exhibits a peak at a value around the
amplitude value of	the sinusoidal variation in (b). The toggling phenomenon manifests
itself as deviations from a pure sinusoidal variation in (b) and
contributes to a width around the peak in the distribution in (c). It is
remarkable that our phenomenological model~(\ref{Eqn:A}) is able to
reproduce  in a rather striking manner the experimental features of the coin
dynamics in (a) and the velocity distribution in (c). Moreover, the
distribution has a tail that is exponential, as seen in panel (c). In
simulations, this is a result of dissipation set by $\gamma$ in the
dynamics~(\ref{Eqn:A}).  Moreover, the amplitude and the phase response as obtained from the simulations are in good agreement with the experimental data (see Fig. \ref{Fig2} (d) and (e)).
As for the coin-trick  problem that was proposed in  the introduction, we  expect the coin  to move with a constant velocity. The magnitude of the velocity would depend on the  coefficient of  friction and the acceleration of the card being pulled out.

\subsection{ A sphere on an orbital shaker }

In the above experiment, the stochasticity in the forcing term arises
from the toggling between the `high-friction' and the `low-friction' state of the coin. In the
second experiment that we describe now, we place a stainless steel ball (a sphere of mass $m$, radius $a$, and moment of inertia $I$) on a plate connected
to an orbital shaker.  The schematic setup of the experimental setup is shown in Fig.~\ref{Fig1}(b). \\

\begin{figure}[t]
    \centering
    \includegraphics[width=.95\linewidth]{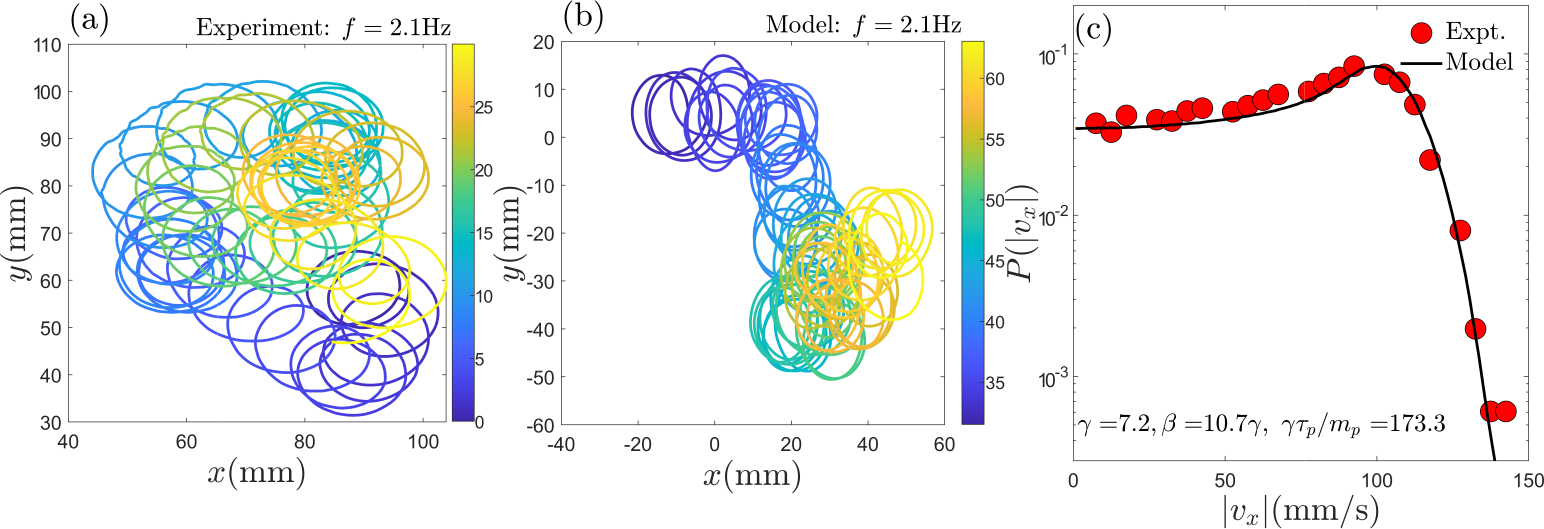}
    \caption{Panel (a) shows the experimental trajectory of a single
        spherical particle on an orbital shaker driven at frequency $f =
        2.1$ Hz, in a reference frame co-moving with the orbital shaker.
        Panel (b) shows  the simulated trajectory based on Eqn.~\ref{Eqn:B}. The colorbar in (a) and (b) indicates time.  Panel (c) shows the probability distribution of the magnitude of
        the $x$-component $v_x$ of the velocity of the center of mass of
        the ball in the co-moving reference frame. The red filled symbols are from the experiment while the black solid line is from the simulated trajectory.
    }
    \label{Fig3}
\end{figure}

Here, the system can toggle between the rolling (without slipping)
and the sliding state. With an orbital shaker, the plate moves in an orientation-preserving manner such that each point on the plate moves on a circle of the same radius $r$: a given point $Q(x_{\rm p},y_{\rm p})$ on the plate moves as
$x_{\rm p}=x_{\rm o}+r\cos (2\pi f t)$, $y_{\rm p}=y_{\rm o}
+r\sin(2\pi f t)$, where $(x_{\rm o},y_{\rm o})$ is the center of the circle
about which the chosen point $Q$ moves. Each point on the plate  has a
center associated with it. The motion of the sphere comes from the force
that is exerted at the point of contact between the sphere and the
moving plate.  This frictional force affecting the velocity
$\mathbf{v}$ of the center of mass of the ball as $m\partial_t \mathbf{v}=-\mu
m g \mathbf{\hat{v}_{\rm p}}$, with $\mu$ the coefficient of friction,
$g$ the acceleration due to gravity, and $\mathbf{v}_p$ being the velocity of
a point on the plate, exerts a torque $\Gamma$ about an axis
parallel to the platform and passing through the center of mass of the
sphere, i.e.,
\begin{equation}
    I\partial_t \mathbf{\omega}_b=-a\hat{k}\times m
    \partial_t\mathbf{v}=a\mu mg
    (\hat{k}\times\mathbf{\hat{v}}_{\rm p})
\end{equation}
where $\mathbf{\omega}_b$ is the angular velocity of the sphere.
An isolated sphere on the plate performs a swirling motion
~~\cite{scherer2000swirling,aumaitre2001segregation,aumaitre_granular_2003}. An example of such a motion for $f=
2.1\,\mathrm{Hz}$ in the reference frame of the moving plate is shown in Fig.~\ref{Fig3}(a). In experiments, the long-time
trajectory of both the position and the velocity of the sphere is not
deterministic. \\

\subsubsection{Experimental results}
\begin{figure*}[t]
    \centering
    \includegraphics[width=.9\linewidth]{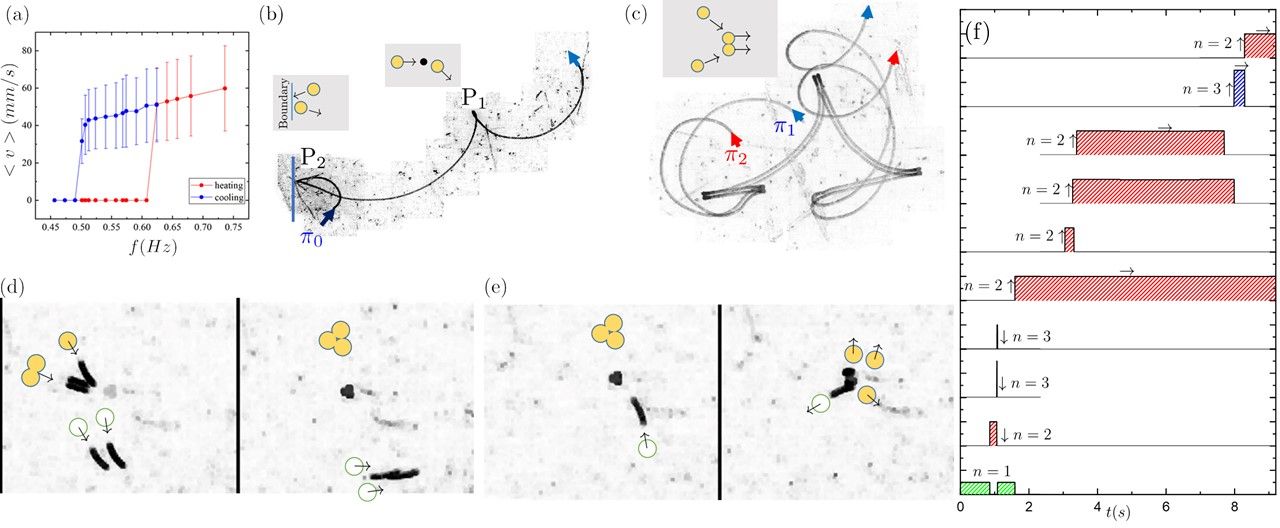}
    \caption{(a) Transitions between stuck to moving state and moving to stuck state occur at different frequencies. (b)  The trajectory of a single sphere shows the two scattering events $P_1$ (particle-substrate) and $P_2$ (particle-wall).  (c) The trajectories $\pi_1$  and $\pi_2$ in the  figure shows the  formation  and destabilization of   dyad like structures.   In the dyad state the trajectories are almost parallel to each other. The figures  (d)  and (e) indicate formation and destabilization of triad (a three particle cluster). Panel (f) Shows the   sequence of formation  and dissolution of different  particle clusters. The size $n$  of these clusters are marked beside each plot. There are 11 particles on the plate and $f=1.9$ $\rm{Hz}$.}
    \label{Fig4}
\end{figure*}

Effectively, this is a two-dimensional
version of the previous problem, wherein the particle-substrate
interaction generates the randomness in the motion. Individually, both the $x$- and the
$y$-components of the motion are approximately sinusoidal in nature. The  abrupt alteration in the trajectory  marked as $P_1$ in Fig. \ref{Fig4} (b) is an example of  such scattering.  Occasional collisions
with the boundary leads to additional randomness in the trajectories. An instance of this kind  of scattering is shown as  $P_2$ in Fig. \ref{Fig4} (b). These scattering events introduce a random phase difference between the two components which
results in the randomization of the  trajectories.
The  probability
distribution of the magnitude of the $x$-component of the velocity, $|v_x|$,
of a rolling sphere for representative value of $f$ is plotted in
Fig.~\ref{Fig3}(c). A similar distribution is seen for $|v_y|$ also. In
the tails, the velocity distribution varies as $P(|\mathbf{v}|)\sim
\exp(-|\mathbf{v}|^\alpha)$, with $\alpha \approx 1$. This is very
different from the well-known Maxwell distribution, in which case $\alpha=2$.   \\

\subsubsection{Model} 	Similar to previous experiment, we postulate that the noise in the	trajectory arises from occasional slipping of the ball with respect
to the plate. These partial slips have a concomitant toggling between
the sliding and the rolling state of the system, and hence, similar to
the previous example, the resulting noise can also be considered
dichotomous. Thus, the Langevin-like equation of motion of the velocity
of the center of	mass of the ball has the form

\begin{equation}
    m \partial_tv_{\sigma}=-\gamma (v_{\sigma}-v_{p \sigma})+\xi(t)(F_{\sigma}(t)+\gamma (v_{\sigma}-v_{p \sigma}))
    \label{Eqn:B}
\end{equation}

with $\sigma =\{x, y\}$. Here, $\xi(t)=0$ and $1$ correspond to sliding and rolling states respectively. It is to be noted that contrary to the coin on the plate problem, here momentum transfer from the plate to the sphere happens in the rolling (`low-friction') state. This is due to the fact in a pure rolling motion (without slipping), the contact point of the sphere is temporarily at rest with respect to the plate. On the other hand, in the sliding state which in this case is a relatively `high-friction' state, the sphere moves in the presence of the damping force.

However, when there are more than one particle in the
system, the inter-particle collisions also contribute to the over all
noise in the trajectories. In the following subsection, we consider the
case of many spheres on an orbital shaker.

\subsection{ Collection of  spheres on an orbital shaker }

In the third experiment, we put $N$ number of balls on the same setup. Care has been taken to ensure that the driven plate is kept as much horizontal as possible which can be seen from the inset in Fig.~\ref{Fig5} (b). It shows the averaged intensity of a stack of 600 images taken at an interval of 1 s. The homogeneous distribution of particle density clearly shows that particles are mostly concentrated at the center of the plate. This also indicates that particle-wall collisions occur rarely as compared to inter particle collisions (wall is indicated by the dotted circular red line).\\

\subsubsection{Experimental Results}
The trajectories shown in Fig.~\ref{Fig4} (b) and (c) are obtained by  averaging a sequence of images, each taken with a time interval of $100$ \rm{ms}.  The micrographs in Fig.~\ref{Fig4} (c) to (d)  highlight  the influence exerted by one particle on the other. When two spheres collide tangentially, two events can occur -  (i) the spheres alter their direction and continue swirling,  or, (ii) they form a bound dyad-like pair; This pair moves together  mostly in a reciprocating  manner,   rolling  back  and  forth in the direction that is perpendicular to the line joining the centers of the two balls. In the direction  that is along the line joining the two centers, the pair  does a  sliding motion. Figure \ref{Fig4}(c) shows two trajectories $\pi_1$  and $\pi_2$ that correspond to a collision and formation of a transiently stable dyad structure. In its dyad form, the two trajectories  move parallel to each other. This structure is short lived and often spontaneously break apart. When it does so, the trajectories $\pi_1$ and $\pi_2$ depart from each other. The destabilization of the dyads can be brought about either by a  particle-substrate  or a  particle-particle scattering event.   In the case of the trajectory shown in  Fig.~\ref{Fig4}(c), it is the particle-substrate scattering that destabilizes the dyad. \\

Occasionally, a third ball bumps into the two-particle pair.  This can either destabilize the dyad  or can form a compact three-particle triad. For spheres in contact,  rolling in the same direction causes shearing of the contact region ~~\cite{kumar2015granular}. Thus,  a triad cannot roll and, hence, it either becomes static or does small sliding motion.  Figure \ref{Fig4}(d) shows a sequence of micrographs that captures the event corresponding to the formation of the  triad.   These micrographs are obtained by  using  exposure time of  $10$ \rm{ms}.  This long exposure  produces a motion-induced dilation effect of the objects. Thus, moving  objects in these micrographs will appear  extended, and its shape tells us about the nature of motion, e.g., a roller's motion will appear as a curved line, and dyads will  appear as  parallel straight lines. For static objects,  motion-induced elongation is absent, and the objects appear as dots.  The leftmost frame  shows a dyad and a nearby rolling isolated sphere. The  other two  particles  in the frame are present to provide a reference.
After collision, the three spheres form a  static triad. Absence of substantial motion for this triad can be inferred from the lack of motion- induced dilation (see the second 	frame of the micrograph in Fig.~\ref{Fig4} (d) and first frame in Fig.~\ref{Fig4} (e)). For any higher-order structure involving more than 3 spheres that can form, the  only available mode of movement is sliding.  So the transitions in these structures are limited to stuck to sliding states. For the values of $f$ reported in this paper, the clusters (dyads, triads and  higher order structures)  are only transiently stable.   Structures involving more than 3 particles are destabilized by particle collisions.  An instance of this destabilization  can be seen in Fig.~\ref{Fig4}(e).  Figure ~\ref{Fig4}(f) shows the corresponding sequence of formation and dissolution of particle clusters for a system of 11 particles that is being driven. The system begins with no clustering, i.e., $n=1$ and then quickly shows clustering. We find that  dyads $n=2$ are the most frequently formed clusters  and that these can be stable for few seconds. Occasionally, we do observe triads $n=3$, but these structures  compared to the dyads are relatively less stable.\\

The very fact that two moving particles can come to a halt after a collision  points  us to instances where the momentum is not conserved. Moreover, the transitions between the rolling, siding and the stuck states are hysteretic in nature, i.e., the frequency  at which a moving particle makes a transition to the stuck state is lower than the corresponding transition from a stuck to a moving state. Figure. \ref{Fig4} (a)  shows this hysteresis  between stuck and  rolling states for a single sphere. \\

\begin{figure}[t]
    \centering
    \includegraphics[width=.95\linewidth]{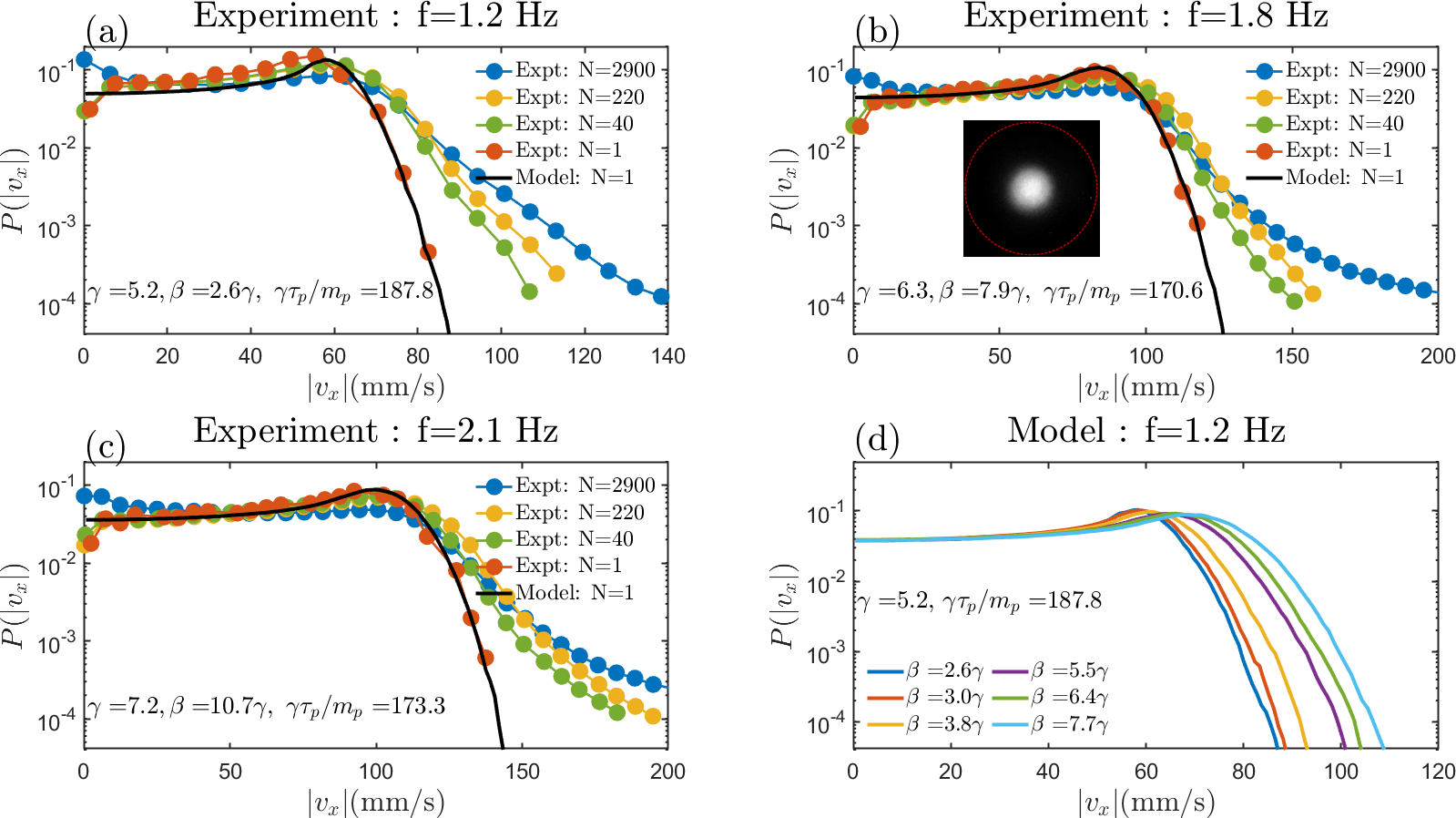}
    \caption{Panel (a), (b) and (c) show the distribution of $|v_x|$ for different number of particles $N$ when the orbital shaker is driven at $f = 1.2$ Hz, $1.8$ Hz and $2.1$ Hz respectively.  The solid black  lines in (a)-(c) are obtained  solving the  phenomenological model (Eqn.~\ref{Eqn:B}) for the respective frequencies. The parameters used for the numerical simulation  are  provided in the figure. 		Inset (b) shows the homogeneous distribution of particles concentrated around the center of the plate away from the wall (red dotted line).  Corresponding distribution from simulation of the phenomenological model (Eqn.~\ref{Eqn:B}) for $f = 1.2$ Hz is shown in panel (d) for different values of $\beta$.
    }
    \label{Fig5}
\end{figure}

With each hysteretic transition, a certain amount of kinetic energy is lost. It would thus be natural to expect higher particle density to result in higher frictional dissipation and hence, lower velocities.  However, contrary to this, we observe in Fig. \ref{Fig5} (a), (b) and (c)  the tail of the velocity distribution $P(|\mathbf{v}|)\sim \exp(-|\mathbf{v}|^\alpha)$, with $\alpha \approx 1$, to increase with the density.

\subsection{Model}
There is as yet no agreement on the value of $\alpha$, with different experiments showing different values, e.g.,  references ~\cite{losert1999velocity,rouyer2000velocity}	 report  $\alpha \approx 3/2$  as an universal parameter, while other experiments ~\cite{kudrolli2000non,van2004velocity,olafsen1999velocity,blair2001velocity} find $\alpha$  to depend on system parameters, where $\alpha$ could take on values close to 2. There
exists multiple models to explain this deviation, e.g., presence of clustering ~\cite{van1998velocity}, non-uniformity of granular temperature ~\cite{nie2000dynamics,puglisi1999kinetic}, nature of noise injected by the  drive  ~\cite{prasad2017velocity,puglisi1999kinetic}. In  most experiments, the strength of the noise is  assumed to be a parameter controlled externally, e.g., energising a shaker more injects more noise to a granular gas, and it is assumed to be independent of the density of the particles. In the present situation, noise is injected when  the particle toggles its state. Thus, with increasing density, the toggling increases in ways that are described in Fig. \ref{Fig4}, and, hence, the strength of the noise increases.

In our model dynamics Eq. (\ref{Eqn:B}), which we now invoke as an effective single-particle dynamics to describe qualitatively the features observed in our experiment, the parameter $\beta$ parameterizes the frequency of this toggling. The model  does well in reproducing the single particle velocity distribution functions   for different values of the drive frequency. The data obtained from  the numerical model are shown as black  lines  in Fig. \ref{Fig5} (a-c). For the many particle case it is mainly the  initial part of the tail  that fits well with the data.

\subsection{The uniqueness of the stationary state}
We  now provide justification  for claiming  that the stationary state obtained  for the experiments mentioned in this paper is qualitatively different from that observed in other driven systems.  There is a large class of driven systems  where the energy injection happens  at the boundary ~\cite{kudrolli1997cluster,falcon1999cluster,opsomer2011phase,noirhomme2018threshold,bougie2002shocks,meerson2004giant,eshuis2007phase,sack2013energy,clement1991fluidization,warr1995fluidization,olafsen1999velocity,kudrolli2000non,blair2001velocity,rouyer2000velocity}, e.g.,  balls on an electromagnetic shaker vibrating vertically or balls in a sinusoidally driven boxes. The height gained by the ball  increases with the increase in  frequency  of the shaker (provided the  peak displacement of the shaker   does not change with the frequency).  This is qualitatively opposite to what we observe for the coin on the moving plate  where coin's displacement reduces with increasing frequency of the moving plate. It may be noted here that for the coin on the plate, the point of energy injection is not spatially separated from  where the energy is dissipated. It is the point of contact with the plate through which the system gains energy through momentum transfer, and it is also the same point through which energy is dissipated via friction. Similarly, the  velocity  distribution obtained for the  spheres on the orbital shaker is also qualitatively different from that obtained in other  experiments  involving  vibrated granular materials.    For most   low density systems,  the  tail of the velocity distribution varies as $P(|\mathbf{v}|) \approx	\exp(-|\mathbf{v}|^\alpha)$. There is a large spread in the value of $\alpha$ reported  in literature ranging from $ 3/2 $  to $2$. In  contrast, we find that for the present method of driving, the tail of the velocity  distribution has an exponential decay, i.e.,  $\alpha \approx 1$. Moreover, the  strength of the  tail grows with increasing density.

\section*{Conclusion}
In conclusion we first summarize  the main experimental findings of the paper,   then enumerate the salient features and the assumptions inherent  in developing the state dependent driving  model of  the systems and  comment on the success and shortcomings of the model.

\paragraph{\textbf{Experiments:}} Across all the three experiments we find the following   essential  features. (i) Abrupt alterations in the dynamics are associated with  toggling of the frictional state. (ii)  The   tail of the  velocity distribution  has an exponential nature. In the context of many-particle systems, this tail gains weight with increasing density.

\paragraph{\textbf{Model:}} We  have constructed a   one-particle theoretical model that captures the essence of the experimental findings. Within our model, we treat the injection of energy into the system to be dependent on its dynamical state.  In writing down the model we have made the following assumptions.
\begin{enumerate}
    \item We  have treated the frictional coupling parameter $\xi$  as a stochastic  variable whose togging in time is taken to be a  Markov process that has  two states; on ($\xi=1$) and  off ($\xi=0$). This Markovian assumption  is a  simplification of the  underlying frictional process that is known to have memory associated with it.
    \item There is a single time scale in the problem given by the constant parameter $\beta$ whose inverse sets the time scale of toggling between the frictional states.
    \item  In absence of a   microscopic  model that establishes the functional dependence between the $\gamma$ and $1/\beta$, we  have chosen both  to linearly vary with   $f-f_c$.
\end{enumerate}
In spite of these assumptions,  we find that our state dependent energy injection model does well in capturing the features  of  the experiments  that are enumerated above. However in terms of exact matching this model leaves room for further improvement for many particle systems, see Fig.~\ref{Fig5}, panels (a)-(d), by, e.g., relaxing the assumptions mentioned above.

Systems with multiple components maintain themselves in a stable state by constantly regulating energy flow and dissipation, e.g., homeostasis in a biological setting is maintained by constant regulation of chemical processes, a  governor in a  combustion engine uses the inertial forces acting on it to limit the fuel injection.  In physical terms, one can think of the system to be consisting of two parts, the body and the environment. The body is the place where the energy is dissipated and to maintain the processes in the body energy has to flow from the environment to the body. The rate of dissipation is a function of the inward energy flux and by controlling this flux, the body maintains a desired stationary state.  Though such self-sustained stationary states are common to biological, chemical and system science settings, there seems to be a gap in realizing this regulation process in a  more prosaic physical setting, particularly in situations where the energy injection and dissipation processes are clearly identified. In this paper, we have shown that for multiple experimental settings that span from a single particle to multiple particles,  a state-dependent energy injection process can lead to a nontrivial stationary state in driven frictional systems.  This stationary state is maintained by continued toggling between the different frictional states of the system. The energy injected to any of these states is a function of the state itself.  We also provide a simple single parameter theoretical description of the various experimental realizations. Our work provides a new paradigm for finding routes to achieve non-equilibrium stationary states. It is only natural that future work in this direction would be to understand the conditions under which the stationary-state is maintained and the processes by which a given stationary state becomes unstable and a new stationary state is arrived at. This could possibly open new avenues to understand modes of failure leading to destabilization in complex systems.
\vskip6pt

\section*{Method}
\textbf{Experiment 1:} A brass coin of diameter $20$ $\rm{mm}$ is placed on a groove on an aluminum plate. The groove has length $300$ $\rm{mm}$ and a thickness slightly larger than the diameter of the coin so that the coin is restricted to move in the $x$ direction only. The plate is kept on top of a shaker executing simple harmonic motion in the $x$ direction, see Fig.~\ref{Fig1}(a). The amplitude of the vibration ($x_{\rm{p0}}$) is $5.4$ $\rm{mm}$. The motion of both the coin and the plate is observed simultaneously at $180$ $\rm{fps}$ from a camera fixed in the laboratory frame for a duration of $120$ $\rm{s}$. The coefficient of static friction measured between the coin and the plate is $\approx 0.42$. The critical frequency below which the coin is completely stuck to the plate is calculated to be $f_c = \frac{1}{2 \pi} \sqrt{\frac{\mu g}{x_{\rm{p0}}}} \approx 4.4$ $\rm{Hz}$. Image processing is done in Matlab. We have done the experiments for other pairs of materials namely Aluminium coin on Aluminium plate, Acrylic coin on Aluminium plate and Acrylic coin on Acrylic plate. The experimental results as well as the results from numerical simulation for these pairs are provided in appendix.

\textbf{Experiment 2:} A single stainless steel ball of diameter $800$ $\mathrm{\mu m}$ is confined on an cylindrical container of diameter $150$ $\mathrm{mm}$ and height $1 $ $\mathrm{mm}$ made of anodized aluminum, with a glass plate used as a lid for the container to enforce a two-dimensional geometry for the problem. The container is placed on a square platform, and the entire assembly along with a camera (used for imaging) is made to perform an orientation-preserving horizontal circular motion (orbital motion). The amplitude of the circular motion ($r$) is $12$ $\rm{mm}$. Images are captured at $110$ $\rm{fps}$ at a resolution of $6.6$ $\rm{MP}$ for a duration of $600$ $\rm{s}$. Image processing is done in Matlab and ImageJ.

\textbf{Experiment 3:} $N$ number of stainless steel balls are placed on the same setup as Experiment 2. Particle tracking is done by using the algorithm mentioned in ~\cite{jaqaman2008robust}.

\appendix
\section{Appendices}

\subsection{Coin on a moving plate : Comparison between experimental and numerical results across  different pairs of materials. }

At the microscopic level, the toggling between the frictional states is determined by the formation and rupture of the contact zone between the particle and the moving plate.  Thus the noise associated with the toggling is closely related to the dynamics of this contact zone (interface). For the contact to break  a critical   yielding stress   of the contact zone needs to be exceeded.   In the framework of rate-state model of friction,
for the contact to break  a critical slip length  $\ell_c$
needs to be reached \cite{dieterich1979modeling}.  Within this model, $\ell_c$ is a linear function of the  frictional coupling strength $\gamma$.

In general the coupling between the plate and the particle is viscoelastic \cite{sharma2008microrheology,kumar2013weak}, and hence the frequency response of $\gamma$ can be  written in terms of the storage modulus ($\mathcal{G}'$) and loss modulus ($\mathcal{G}''$) as,
\begin{equation}
    \gamma=\ell\left(\frac{\mathcal{G}''-i \mathcal{G}'}{\omega}\right).
\end{equation}
Here $\omega =2\pi f$, and $\ell$ is the characteristic length scale of the coin. Within the framework of Maxwell’s model of viscoelasticity,

\begin{eqnarray}
    \mathcal{G}'&=& G+\left(\frac{\eta_0}{\tau_p}\right) \left(\frac{\omega^2\tau_p^2}{1+\omega^2\tau_p^2}\right) \\
    \mathcal{G}''&=& \left(\frac{\eta_0}{\tau_p}\right)\left(\frac{\omega\tau_p}{1+\omega^2\tau_p^2}\right)
\end{eqnarray}
Here $\eta_0$  and $G$ are the  zero-frequency viscosity  and rigidity, respectively,  and $\tau_p$ is the only relaxation time in the problem.

The following equation  describes the motion of the particle (see the main paper for details).
\begin{equation}
    m \partial_tv=-\gamma (v-v_p)+\xi(t)(F(t)+\gamma (v-v_p)).
    \label{Eqn:A}
\end{equation}

For $x_p =x_{p0}e^{i\omega t}$  and   $x=x_{c0}e^{i(\omega t +\phi)}$ we have

\begin{eqnarray}
    x_{c0}e^{i\phi}&=& -i x_{po}\left( \frac{\left(\mathcal{G}''-i \mathcal{G}'\right) \left( 1-\langle \xi \rangle \right)\tau_p \ell + \left(\langle \xi \rangle m_p \omega \right) }{m  \tau_p\omega^2 -i  \left(\mathcal{G}''-i \mathcal{G}' \right) (1- \langle \xi \rangle ) \tau_p \ell}\right).
\end{eqnarray}

Here $\phi$ is the phase difference between the plate and the coin and $\langle \xi \rangle$ which is a function of the toggling rate $\beta$ is the time averaged value of the dichotomous noise $\xi(t)$.

For values of $\omega\tau_p \ll 1$,
\begin{eqnarray}
    Im(\gamma)  = \frac{\ell\mathcal{G}'}{ \omega}&=&\ell\left( \frac{G}{\omega}+\eta_0 \omega\tau_p \right)\\
    Re(\gamma) = \frac{\ell\mathcal{G}''}{\omega}&=& \eta_0\ell
\end{eqnarray}
In the limit  of $\omega \rightarrow 0$,
we  obtain a frequency weakening response of the frictional coupling $\gamma$, i.e.,
$\mathcal{G}' = G$ and $\mathcal{G}''$ = 0.
\begin{eqnarray}
    \gamma &=& -\frac{i G\ell}{\omega}\\
    \tan \phi &\approx &\frac{\langle \xi \rangle m_p \omega}{\ell G\tau_p(1- \langle \xi \rangle ) }     \\
    x_{co} &=& \frac{\sqrt{( G (1- \langle \xi \rangle )\tau_p \ell)^2+( \langle \xi \rangle m_p \omega)^2}}{m\tau_p\omega^2-  G(1- \langle \xi \rangle ) \tau_p \ell}
\end{eqnarray}
In this limit  
$\tan \phi \propto \omega \propto f $
and $x_{co} \propto 1/\omega \propto 1/f$. These observations are consistent with our experimental observations.


The stochasticity associated with the toggling process   has its microscopic origin in the  variation of the  instantaneous value of the  length $\ell(t)$ at which the slip is initiated. This is closely related to the  yielding process of the interface and hence the strength of the noise associated with this process should be proportional to the mechanical properties of the interface, thus $1/\beta$  should be a function of the frictional coupling that is described by  $\gamma$.

\begin{figure}[t]
    \centering
    \includegraphics[width=.99\linewidth]{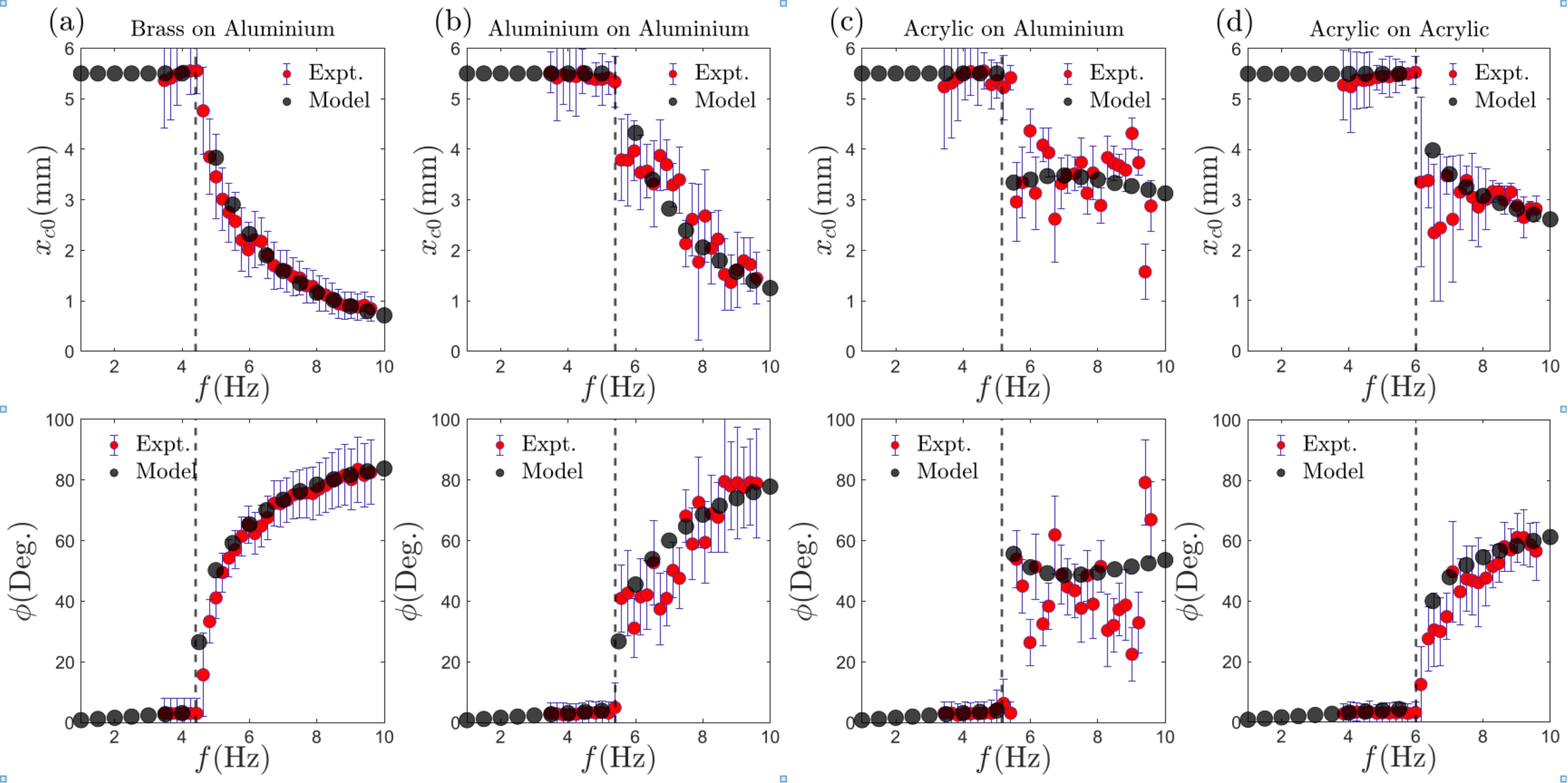}
    \caption{The figure compares the values of the   amplitude $x_{c0}$(top panel) and the phase lag $\phi$ (bottom panel)  obtained from the experiments to that obtained  from the model.  The comparison is done for four different pairs of materials. These pairs are mentioned in the figure. 
        The parameters used in the simulations are tabulated in Table. S1.   }
    \label{fig:Suppl_1}
\end{figure}

\begin{table}[b]
    \centering
    \begin{tabular}{llllllll}
        \cline{1-8}
        Coin on Plate           & $f_c$  & $\kappa_0$ & $\gamma_0$ & $\gamma_s$ & $\beta_0 $ & $\beta_s$ & $\mu_s$ \\
        &  (Hz) & $ $ & (kg/s) & $ (kg/s) $ & (Hz)  & (Hz)  \\

        \cline{1-8}
        Brass on Aluminium      & 4.4        & 12         & 2.2        & 10.7       & 238    & 1562   & 0.21 \\
        Aluminium on  Aluminium & 5.4        & 12         & 4          & 14         & 238    & 1562   & 0.45 \\
        Acrylic on Aluminium    & 5.15       & 55         & 55         & 7          & 222    & 2000     & 0.38 \\
        Acrylic on Acrylic      & 6          & 55         & 55         & 3          & 1250    &
        500   & 0.66 \\
        \cline{1-8}
    \end{tabular}
    \label{Table:tabl_param}
    \caption{Tabulation of the parameters used for the numerical simulation of the model which is shown in Fig. \ref{fig:Suppl_1}. Here $\kappa_0=(\gamma \tau_p)/(m_p)$. We use the term Acrylic to describe polymethyl methacrylate (PMMA).}
\end{table}

In absence of a   microscopic  model that establishes the functional dependence between the two, we  have chosen both $\gamma$ and $1/\beta $ to linearly vary with $\Delta f$. Here $\Delta f = |(f-f_c)|/f_c$ is the reduced frequency of the drive and $f_c$ denotes the critical drive frequency beyond which the coin begins to slip.   The choice of this linear variation was necessitated to account for the strain rate induced weakening of the frictional coupling.  We take the following  functional forms  of the parameters $\gamma$ and  $1/\beta$ to describe their dependence  on the drive frequency $f$.
\begin{eqnarray}
    \gamma &=& \gamma_s+ \frac{\gamma_0}{\Delta f}\\
    \frac{1}{\beta} &=&  \frac{1}{\beta_s}+\frac{\Delta f}{\beta_0}
\end{eqnarray}

Moreover, the dimensionless parameter
\[ \kappa_0 = (\gamma \tau_p)/ m_p\]
was taken to be a constant   whose value depends on the pair of the material involved in the frictional process. The  comparison between the   results obtained from the simulation and the experiments across different pairs of coin and the plate materials  is  provided in Fig. \ref{fig:Suppl_1}. These results also highlight the contrasting frictional response of  pairs involving hard (metals, $G_{\mathrm{Brass}}\sim 40$ Gpa, $G_{\mathrm{Al}}\sim 20$ GPa) materials and those involving softer (polymethyl methacrylate  (Acrylic), $G_{\mathrm{Acr.}}\sim 2$ GPa) materials. Hard materials with larger values of elastic constant $G$  have larger critical  slip lengths and hence are less prone to frictional instabilities, these materials  exhibit a  continuous type of transition across $f_c$. In contrast, the pairs involving acrylic (a softer material that  lower rigidity modulus $G$ and hence small critical slip length $\ell_c$) are more prone to frictional instabilities  and hence exhibit an abrupt jump across the transition. These features are  captured well within the model. The comparison of the experimental findings with that obtained from the  model are provided in  Fig. \ref{fig:Suppl_1} for different pairs of materials.

\bigskip

\subsection{Predictions from the Coloumb Model}
\begin{figure}[t]
    \centering
    \includegraphics[width=0.95\linewidth]{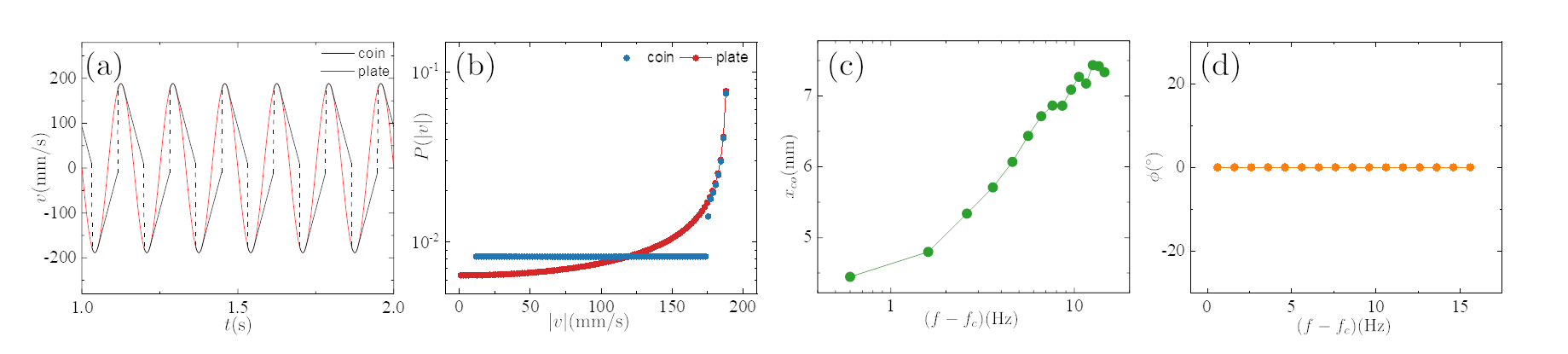}
    \caption{A coin on an oscillating plate:  The panel (a) shows  the  trace of the velocity of the coin (black) and a point fixed on the plate (red) at driving frequency $f=6$ Hz as obtained from the Coloumb model of friction.     The probability distribution of the magnitude of the velocity of the coin in the laboratory frame is plotted in panel (b) at the same value of $f$. The measured time-averaged amplitude of the displacement of the coin ($x_{\rm c0}$) is shown in panel (c) as a function of the driving frequency $f$.  Panel (d) shows the phase difference $\phi$ of the coin with respect to the drive as a function of $f$.
    }
    \label{fig:coloumb}
\end{figure}

In performing the simulations for the Coulomb friction model \cite{persson2013sliding}  we have assumed $\mu_s=\mu_k=\mu=0.42$.  The  simulated velocity profile  is shown in Fig.~\ref{fig:coloumb}(a). It can be noted that there are discontinuous jumps in the velocity plot marked as dotted lines. This happens because of the presence of the velocity gap in the coulomb friction model associated with the  transition of the coin  from the slip state to the stuck state as pointed out in section on ``Dry friction models and their limitations'' of the main paper. The corresponding velocity distribution of the coin is plotted in Fig.~\ref{fig:coloumb}(b) (blue dots). As a reference the velocity distribution of the plate is also plotted alongside (red dots). Since the velocity of the coin is linear in time in the slip state, it corresponds to the constant part of the distribution function in Fig.~\ref{fig:coloumb}(b). Whereas, in the stuck state the velocity of the coin follows that of the plate, which corresponds to the part which follows the increasing part of the distribution. A small part of the lower velocity range is absent in the velocity distribution of the coin which falls in  the velocity gap mentioned above. The amplitude and phase difference of the coin with respect to the plate are shown in Fig.~\ref{fig:coloumb}(c) and (d) respectively. Clearly they do not agree with the experimental data.\\
\bibliographystyle{plain} 
\bibliography{velo_dis} 

\enlargethispage{20pt}


    Shamik Gupta acknowledges support from the Science and Engineering Research Board (SERB), India under SERB-TARE scheme Grant No. TAR/2018/000023, SERB-MATRICS scheme Grant No. MTR/2019/000560, and SERB-CRG scheme Grant No. CRG/2020/000596.

The authors Soumen Das and Shankar Ghosh thank Prajwal Panda, Anit Sane and  Soham Bhattacharya for their help in the experiments.



\providecommand{\newblock}{}

\end{document}